\documentclass[12pt,preprint]{aastex}
\usepackage{txfonts}
\usepackage{bbm}
\usepackage{amssymb}

\usepackage{booktabs}
\usepackage{graphicx}

\begin{document}
\title{Buoyant bubbles in intracluster gas: effects of magnetic fields
and anisotropic viscosity}

\shorttitle{3D MHD simulations of bubbles in clusters}

\shortauthors{Dong \& Stone}

\author{Ruobing Dong and James M. Stone}

\affil{Department of Astrophysical Sciences, Princeton University,
Princeton NJ 08544\\}

\begin{abstract}

Recent observations by {\it Chandra} and {\it XMM-Newton} indicate
there are complex structures at the cores of galaxy clusters, such
as cavities and filaments.  One plausible model for the formation of
such structures is the interaction of radio jets with the
intracluster medium (ICM). To investigate this idea, we use
three-dimensional magnetohydrodynamic simulations including
anisotropic (Braginskii) viscosity to study the effect of magnetic
fields on the evolution and morphology of buoyant bubbles in the
ICM.  We investigate a range of different initial magnetic field
geometries and strengths, and study the resulting x-ray surface
brightness distribution for comparison to observed clusters.
Magnetic tension forces and viscous transport along field lines tend
to suppress instabilities parallel, but not perpendicular, to field
lines. Thus, the evolution of the bubble depends strongly on the
initial field geometry.  We find toroidal field loops initially
confined to the interior of the bubble are best able reproduce the
observed cavity structures.

\end{abstract}


\section{INTRODUCTION}

In recent years, $Chandra$ observations of galaxy clusters have
revealed complex structures such as cavities and filaments in the
x-ray surface brightness distribution.  The most prominent examples
include Perseus (Fabian et al. 2000), Cygnus A (Carilli et al.
1994), Virgo-A (Young et al. 2002) and Hydra A (McNamara et al.
2000); a systematic study of 16 clusters is given by Birzan et al.
(2004), and an extensive survey of 64 X-ray cavities in clusters,
groups and elliptical galaxies is given by Diehl et al. 2008.
Typically x-ray cavities are located at about 20 kpc (projected)
from the cluster center, and have a radius of about 10 kpc.
Interestingly, most of these systems harbor central galaxies with
radio luminosities between $2\times 10^{38}$ and $7\times 10^{44}$
ergs s$^{-1}$. Moreover, $Chandra$ observations confirm that these
clusters have an inwardly decreasing temperature profile, and
therefore are so-called ``cool core" clusters (Birzan et al. 2004).

One interpretation of the observations is that such structures are
formed by the interaction of radio jets produced by a supermassive
black hole (SMBH) in the central radio galaxy with the intracluster
plasma.  Such interaction will produced shocked jet material that is
highly over-pressurized with respect to the ambient intracluster
medium (ICM).  As it expands to achieve pressure equilibrium, it
will form a low density, high entropy cavity -- that is a ``bubble".
In the density stratified ICM, the bubble will be buoyant, and as it
rises it evolves into the observed cavity-like structures.  The
feedback of radio jets on the ICM through the dynamics of
pressure-inflated bubbles might be an important source of heating of
the x-ray emitting plasma (Ruszkowski et al.2004; Fabian et al.
2005), and contribute to the prevention of the cooling catastrophe
(Fabian 1994).

In order for this interpretation to be correct, rising bubbles in
the ICM must remain coherent, and avoid being shredded by
Rayleigh-Taylor instability (RTI), Richtmyer-Meshkov instability
(RMI) and Kelvin-Helmholtz instability (KHI).  A variety of authors
have used numerical simulations to study the hydrodynamics of
buoyant bubbles in two-dimensions (e.g., Bruggen \& Kaiser 2001,
2002; Churazov et al. 2001; Bruggen 2003; Reynolds et al. 2002;
Sternberg et al. 2007), and three-dimensions (e.g., Basson \&
Alexander 2003; Quilis et al. 2001; Omma et al. 2004; Gardini 2007;
Pavlovski et al. 2008). Pizzolato \& Soker (2006) have shown that in
the early evolution, when the AGN jet is still inflating the bubble,
it is stable to RTI, and Sternberg \& Soker (2009a) studied the
inflation of the bubbles by highly subrelativistic massive jets, and
gave an example on a galaxy cluster (Sternberg \& Soker, 2009b),
while Fabian et al. (2005) have shown that the heating and cooling
rates associated with weak shocks, viscosity and thermal conduction
vary with temperature and radius in the ICM. Other recent
three-dimensional hydrodynamic simulations include Bruggen et al.
(2009) and Scannapieco \& Bruggen (2008). A general result of all
such studies is that purely inviscid hydrodynamical evolution
generally can not produce the observed morphology of the cavities,
nor reproduce the heating required to explain the observed
temperature profile in the clusters.  This suggests that additional
physics is important to the dynamics, (Gardini 2007; Reynolds et al.
2005; Diehl et al. 2008): possible candidates include turbulent
motions in the ICM, magnetic fields, viscosity and thermal
conduction.

Since the hot gas in clusters of galaxies is likely magnetized
(Carilli \& Taylor 2002; Taylor et al. 2002), magnetohydrodynamic
(MHD) effects could be important for the evolution of buoyant
bubbles.  Bruggen \& Kaiser (2001) studied rising bubbles in
two-dimensional MHD in spherical geometry to investigate the effect
of magnetic fields on the RTI and KHI.  A larger parameter survey,
including various field geometries and strengths, and various
profiles for the atmosphere, was presented by Robinson et al. (2004)
for a two-dimensional planar geometry. Jones \& De Young (2005)
extended this work using spatially varying fields and internal
fields generated self-consistently as the bubble is inflated. More
recently, O'Neill et al. (2009) further extended the work of Jones
\& De Young (2005) to three-dimensional MHD simulations. Nakamura et
al. (2006, 2007) studied the origin of magnetized bubbles. In their
three-dimensional MHD simulations the SMBH injects poloidal and
toroidal magnetic field into space and inflates the bubbles. Other
three-dimensional MHD simulations include Ruszkowski et al. (2007),
which focus on fossil bubbles in the presence of tangled magnetic
fields; Liu et al. (2008), which presented long-term simulations of
magnetized bubbles, and Dursi \$ Pfrommer (2008), showing bubbles
may develop a dynamically important magnetic sheath around
themselves, which can suppresses instabilities at the surface of the
bubbles. All of these authors confirm that a strong magnetic field
parallel to the surface of the bubble can stabilize both KHI and RTI
during the evolution, as expected from the linear stability analysis
(Chandrasekhar 1961; Shore 1992). Fully three-dimensional studies
are important for this problem, however, because magnetic fields
only suppress RTI for modes parallel to the field; they have no
effect on interchange modes perpendicular to the field. Thus, fully
three dimensional studies of the magnetic RTI (Stone \& Gardiner
2007a; b) have shown rapid growth of unstable fingers even in the
case of strong fields.

Viscous effects may also be important to the dynamics
(Fabian et al. 2003), since
the Reynolds number in the intracluster gas is so low.  Using the
standard Spitzer (1962) estimate for the coefficient of viscosity
leads to an estimate of the Reynolds number of
\begin{equation}
Re=62\left(\frac{U}{390\ \rm {km\ s}^{-1}}\right)\left(\frac{L}{20\
\rm{kpc}}\right)\left(\frac{\rm{kT}}{5\
\rm{keV}}\right)^{-2.5}\left(\frac{n}{0.03\
\rm{cm}^{-3}}\right)\left(\frac{\ln\Lambda}{30}\right)
\end{equation}
where $U$ is the typical velocity of motions (scaled to one half the
adiabatic sound speed for hydrogenic plasma at $kT=5$ keV), and $L$ is a
typical length scale for the bubble (Reynolds et al. 2005).  Ruszkowski et
al. (2004) studied viscous dissipation of the gas motions and sound waves
generated by buoyant bubbles, and the implications for heating of the ICM.
Reynolds et al. (2005) studied the effect of viscosity on the morphology
and evolution of bubbles, including comparisons between the synthetic
x-ray surface brightness of bubbles in their hydrodynamic simulations
with observations.  These authors pointed out that a coefficient of
viscosity that is only 25 percent of the Spitzer value can quench the
development of instabilities and maintaining the integrity of the bubble.

However, in a weakly-collisional plasma such as the ICM, microscopic
transport processes such as viscosity and thermal conduction will be
anisotropic when the collision mean free path of ions or electrons is
much larger than their Larmor radius.  For example, momentum transport
by viscosity (which is mediated primarily by ion-ion collisions) will
be anisotropic when
\begin{equation}
\epsilon \equiv (\omega_{c,i} \tau_{c,i}) ^{-1} \ll 1
\end{equation}
where $\omega_{c,i}$ is the ion cyclotron frequency and $\tau_{c,i}$
is the ion-ion mean collision time.   For a hydrogenic plasma, this
condition can be written as
\begin{equation}
\omega_{c,i} \tau_{c,i}=(\frac{1.09\times 10^5}{n})\frac{T_4^{1.5}
B_{\mu G}}{ln\Lambda}
\end{equation}
where $n$ is the proton density in cm$^{-3}$, $T_4$ is the kinetic
temperature in units of $10^4$K, $B_{\mu \mathrm{G}}$ is the
magnetic field in microgauss.  In the ICM, typical temperatures are
$T \sim 10^{7}$~K and typical densities are $n \sim 10^{-3}-10^{-2}$
cm$^{-3}$ (Peterson \& Fabian 2006). The magnetic field strength is
roughly $B \sim 1-10$~$\mu$G at the center of a typical cluster, and
0.1-1 $\mu$G in the outer regions.  For these values, clearly
$\epsilon \ll 1$, and viscous transport will be anisotropic.  In
addition, we note that $\beta = (8\pi p)/(B^2)$, the ratio of the
gas to magnetic pressure, is large for the conditions in a typical
cluster ( $\beta\approx 200 - 2000$ in the central regions) (Parrish
et al. 2008), which is an indication that the magnetic field is
dynamically unimportant.  However, due to anisotropic transport
effects, the existence of even a weak magnetic field will still
dramatically influence the physical transport processes.

Recently, the effect of anisotropic transport processes in weakly
collisional plasmas has been studied both theoretically (Balbus
2004; Islam \& Balbus 2005; Lyutikov 2007, 2008) and numerically
(Sharma et al. 2006; Parrish \& Quataert 2008; Quataert 2008;
Parrish et al. 2008,2009) in a number of astrophysical systems.  In
the context of clusters, Balbus (2000) has shown that anisotropic
thermal conduction produces {\em qualitatively} new effects, such as
modifying the convective stability criterion (the magnetothermal
instability or MTI).  Parrish \& Stone (2005; 2007a; b) and Parrish
et al. (2008) have studied the nonlinear saturation of the MTI,
while Parrish et al. (2008) investigated the effect of the MTI on
the evolution of the temperature profile in the ICM. Quataert (2008)
pointed out that in the presence of a background heat flux,
anisotropic thermal conduction will drive a new buoyancy instability
(the heat flux buoyancy instability, or HBI) when the temperature
decreases in the direction of gravity, and Parrish \& Quataert
(2008), Sharma et al. (2009), Parrish et al. (2009) and Bogdanovic
et al. (2009) have studied the HBI using three-dimensional MHD
simulations.  Lyutikov (2007, 2008) has emphasized the importance of
anisotropic transport in the ICM, arguing that it determines the
dissipative properties, the stability of embedded structures, and
the transition to turbulence in the ICM. Thus, simulations of the
dynamics of the ICM using isotropic versus anisotropic transport
(viscosity or thermal conduction) can produce quite different
results.

The goal of this work is to study the effect of anisotropic
viscosity and magnetic fields on the dynamics of buoyant bubbles in
the ICM. Although other physical processes, such as turbulence and
cosmic rays generated by merging substructures, or thermal
conduction, could be important for interpreting the observations of
structures in the ICM, we will not include such processes here.  Our
goal is simply to investigate the role of magnetic fields and
anisotropic viscosity on the RTI and KHI in rising bubbles, using
different field strengths and directions, and to compare the
morphology of bubbles computed with realistic values of the
parameters with the observations.  Our study can be considered as an
extension of previous MHD simulations that did not include
viscosity, or previous viscous simulations that did not include MHD.

This paper is organized as follows. In \S2 we review the basic physics in
the simulations, and we describe the numerical methods used in this work.
In \S3 we present results from our simulations, including the evolution
of the density profile and simulated x-ray images.  In \S4 we discuss
the implications of our results for models of AGN feedback and heating,
and in \S5 we present our conclusions.

\section{Method}

We solve the equations of MHD, which can be written in conservative form as
\begin{equation}
\frac{\partial \rho}{\partial t} + {\bf\nabla\cdot} (\rho{\bf v}) =
0,
\label{eq:cons_mass} \\
\end{equation}
\begin{equation}
\frac{\partial (\rho {\bf v})}{\partial t} + {\bf\nabla\cdot}
\left(\rho{\bf vv} - \frac{\bf BB}{4\pi} + P + \frac{\bf B^2}{8\pi}
- \bar{\bf \sigma} \right) = - \rho {\bf \nabla}\Phi,
\label{eq:cons_momentum} \\
\end{equation}
\begin{equation}
\frac{\partial(\rho E)}{\partial t} + {\bf\nabla\cdot} \left(\rho E
\bf v + \left[ p+\frac{\bf B^2}{8\pi}\right]\bf v- \frac {(\bf v \cdot \bf B)\bf
B}{4\pi} - \bf v\cdot\bar{\bf \sigma}\right) =
\rho {\bf v}\cdot{\bf \nabla}\Phi,
\label{eq:cons_energy}\\
\end{equation}
\begin{equation}
\frac{\partial {\bf B}}{\partial t} - {\bf\nabla} \times \left({\bf
v} \times {\bf B}\right) = 0, \label{eq:induction}
\end{equation}
where $\Phi$ is a fixed gravitational potential due to the dark
matter, $E$ is the total energy (sum of internal, kinetic and
magnetic energies, but excluding the gravitational potential
energy), and ${\bf \sigma}$ is the viscous stress tensor.  The other
symbols have their usual meaning. We have written these equations in
units such that the magnetic permeability $\mu_{B}=1$.  In a
magnetized plasma, the anisotropic viscous stress tensor can be
written as (Balbus 2004),
\begin{equation}
\sigma_{ij} = \mu \left(\frac{B_m B_k}{\bf B^2} \partial_k v_m -
\frac{\nabla\cdot \bf v}{3}\right) \left(\frac{B_i B_j}{\bf B^2} -
 \delta_{ij}\right)
\end{equation}
where summation over repeated indices is implied.  We fix the coefficient of
viscosity $\mu$ to be a constant parameter for all the simulations
presented here.

The effects of the dark matter on the dynamics
is strictly through gravitational forces represented by a fixed potential.
Following Reynolds et al. (2005), we assume the dark matter potential
is given by
\begin{equation}
\Phi = \frac{0.75}{\gamma}\log(1+r^2)
\end{equation}
We assume the ICM forms an isothermal hydrostatic equilibrium in this
potential, so that $\rho (r) = \rho_0[1+(\frac{r}{r_0})^2]^{-0.75}$
and $P = c_s^{2} \rho$, where we choose units of mass, length, and time
such that $\rho_0=1$, $r_0=1$, and $c_s=1$.

At the start of the calculation, we introduce a bubble into this
atmosphere as a spherical region offset from the center of the dark matter
potential a distance $R=0.3$, with density $\rho_{b}=0.01(1+\delta)$ and
radius $r_{b}=0.25$.  We introduce perturbations to the density in the
bubble through $\delta$, a random number chosen independently for each
computational cell contained within the bubble, with $| \delta | \leq
0.005$.  We find such perturbations are necessary to break the symmetry
of the resulting evolution, otherwise our numerical scheme will preserve
such symmetries exactly.  Initially the bubble is in pressure equilibrium
with its surroundings.  The evolution of the bubble is computed using an
adiabatic equation of state with $\gamma=5/3$.  Note we do not attempt
to model the formation and inflation of the bubble via the interaction
of a radio jet with the ICM.  Instead, and following previous authors,
we simply introduce a buoyant bubble into a hydrostatic atmosphere and
follow the resulting dynamics.

Our calculations use Athena (Stone et al.  2008), a recently
developed Godunov scheme for astrophysical MHD.  The mathematical
foundations of the algorithms are documented in Gardiner \& Stone
(2005; 2008). The directionally unsplit CTU (corner transport
upwind) integration scheme, combined with the Roe Riemann solver,
are used throughout. Extensive tests of the algorithms are presented
in Stone et al. (2008). The gravitational source terms due to the
dark matter potential are added to the interface reconstruction,
transverse flux gradient corrections, and final corrector update in
the CTU integration algorithm directly to preserve second order
accuracy, and in a way that conserves the total energy (including
gravitational potential energy) exactly.  Anisotropic viscosity is
added through operator splitting.  In order to conserve total
momentum exactly, we difference the viscous fluxes at each cell
interface, which requires that the components of ${\bf \sigma}$ be
centered at the appropriate cell faces.  We use TVD (total variation
diminishing) slope limiting to construct the velocity differences
when computing the stress, using a method that is exactly analogous
to computing the heat fluxes with anisotropic conduction (Sharma \&
Hammett 2007). This step is crucial to preventing unphysical
momentum fluxes.  The details of our algorithm are presented in the
Appendix in Stone \& Balbus (2009).

Our simulations use a Cartesian grid that spans a domain $-1 \leq x
\leq 1$, $0 \leq y \leq 3$, and $-1 \leq z \leq 1$.  The origin of
the dark matter potential $r=0$ is located at $(x,y,z) = (0,0,0)$,
and the center of the bubble is located initially along the $y-$axis
at $(x,y,z)=(0,0.3,0)$.  The bubble rises in the $y-$direction,
hereafter we refer to this direction as the ``vertical".  Our
typical numerical resolution is $256\times384\times256$, although we
present a convergence study in which we both double and halve this
resolution in each direction.  We use reflecting boundary condition
at $y=0$ and outflow boundary conditions everywhere else.  At
outflow boundaries, we project all quantities with zero slope,
except the pressure, which we compute from integrating the equation
of hydrostatic equilibrium into the ghost zones using the pressure
in the last active cell as an initial condition.  We find this step
maintains hydrostatic equilibrium near the boundaries of the domain
quite well. Even though we allow material to flow freely through the
boundaries, we find that by the end of each simulation the mass lost
is less than $5\times 10^{-4}$ of the initial value.  Due to our
outflowing boundary conditions, the total energy of the gas
(including the gravitational potential energy) in our simulations is
not conserved, although we find that the fractional change in the
total energy in the domain is small, less than $10^{-3}$. Lack of
strict conservation of energy complicates the measurement of the
work done on the gas by rising bubbles, we examine this issue in
more detail in \S4.2.

We study three different initial magnetic field configurations, including
(1) a uniform horizontal field ${\bf B} = (B_0,0,0)$, (2) a uniform
vertical field ${\bf B} = (0,B_0,0)$, and (3) a uniform toroidal field of
strength $B_0$
located only inside the bubble.  In the first two cases the field is
located throughout the domain (within both the bubble and the atmosphere).
In the toroidal field case, the field lines form
concentric loops inside the bubble, and there is
no field outside the bubble.  In this case the field is initialized from a
vector potential via ${\bf B} = \nabla \times \bf A$, where the components of
${\bf A}$ are defined at cell edges, with the only
non-zero component
\begin{equation}
A_y = B_0 (\sqrt{0.25^2-(y-0.3)^2} - \sqrt{x^2+z^2}),
\end{equation}
This procedure
guarantees the initial field satisfies the divergence-free constraint
exactly.  For each field geometry, we study two different field strengths:
a weak field given by $\beta = 2\rho_0 c_s^{2}/B_0 =
1.2\times 10^6$, and a strong field given by $\beta=480$.

We can relate the units used in the simulations to quantities in real
clusters by adopting fiducial values for length, mass and
time. For example, if we fix the unit of length used in the code to
20~kpc, the unit of mass to be given by the density $\rho_0=1$ being
equivalent to a proton number density of 0.03~cm$^{-3}$, and the
unit of time to be fixed by the adiabatic sound speed $c_s=1$ being
equivalent to 800 km s$^{-1}$, then our simulation domain spans a region
$40\times40\times60$ kpc, our typical numerical resolution is about
0.16~kpc, and the sound crossing time is about 25~Myr.  Furthermore,
the weak field case corresponds to a magnetic field strength of 0.05 -
0.10 $\mu$G, while the strong field case about 1 - 5 $\mu$G.

\section{Results}

We present the results from 11 different simulations (although we
have run many more to test our methods and survey parameters). For
those simulations that include viscosity, we fix the coefficient of
viscosity $\mu$ to give a Reynolds number $Re = c_s L\rho_0 /2\mu =
50$ (using $c_s/2$ as the characteristic velocity in the problem),
where in the code units $c_s = L = \rho = 1$.  We report results
from two purely hydrodynamic simulations performed as control
experiments for comparison to the MHD cases.  One of these
simulations is inviscid, and one has been run with an isotropic
Navier-Stokes viscosity with a Reynolds number of 50.  The remaining
9 simulations correspond to each of the three magnetic field
geometries (horizontal, vertical, and toroidal) run with both a weak
and strong field with anisotropic viscosity, and a strong field with
no viscosity.  We have confirmed that simulations run with a weak
field and no viscosity are nearly identical to the inviscid
hydrodynamic case, and thus such models will not be discussed
further. We have run the strong field cases both with and without
viscosity in order to isolate the effects of the magnetic field
(MHD) from the effects of the anisotropic viscosity.  Table
\ref{table:property} summarizes the properties of all the
simulations presented here.

\begin{deluxetable}{cccc}
\tabletypesize{\scriptsize} \tablewidth{0pc} \tablecaption{Property
of the simulations presented in this paper} \tablehead{
\colhead{Label} & \colhead{Re} & \colhead{$\beta$} & \colhead{Field
direction}} \startdata
$H1$ & - & - & - \\
$H2$ & 50 & - & - \\
$X1$ & 50 & $1.2\times 10^6$ & horizontal \\
$X2$ & -  & 480 & horizontal \\
$X3$ & 50 & 480 & horizontal \\
$Y1$ & 50 & $1.2\times 10^6$ & vertical \\
$Y2$ & -  & 480 & vertical  \\
$Y3$ & 50 & 480 & vertical  \\
$T1$ & 50 & $1.2\times 10^6$ & toroidal \\
$T2$ & -  & 480 & toroidal  \\
$T3$ & 50 & 480 & toroidal  \\
\enddata
\label{table:property}
\end{deluxetable}

\subsection{Convergence test}

\begin{figure}
  \includegraphics[width=160mm]{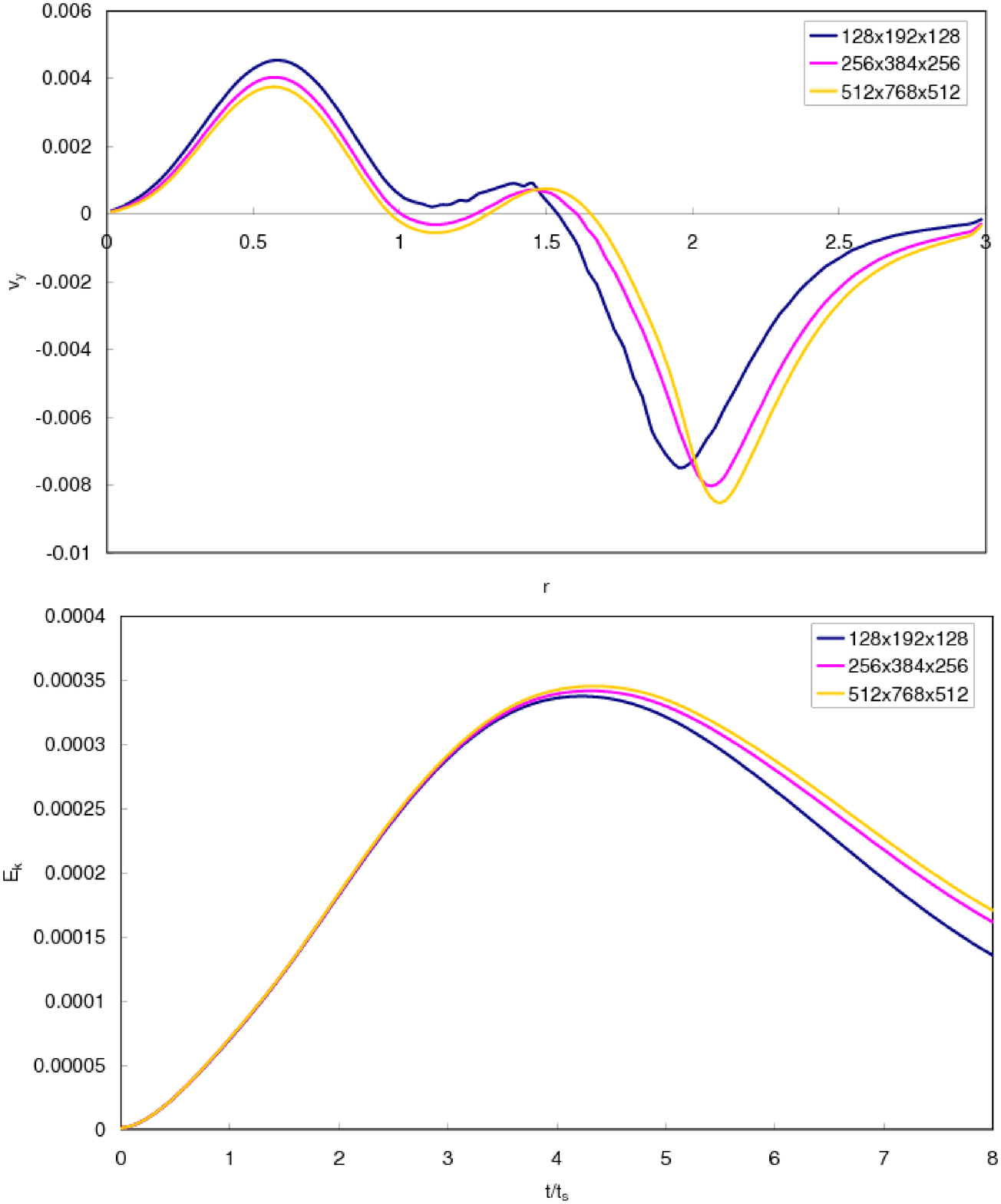}\\
  \caption{({\em Top.}) Radial profile of the angle-averaged vertical velocity
at $t=8$ in run H2 computed with three different numerical
resolutions. ({\em Bottom.}) Volume averaged kinetic energy as a
function of time for run H2 at three different resolutions.}
  \label{ns_con_c}
\end{figure}

It is important to confirm that our standard numerical resolution,
$256\times384\times256$, is sufficient to resolve the flow.   Since
we are studying a relatively low Reynolds number ($Re = 50$), we
expect that achieving numerical convergence will not be difficult.
To study numerical convergence we have run model H2 at three
different resolutions: $128\times192\times128$ (low resolution),
$256\times384\times256$ (standard resolution), and
$512\times768\times512$ (high resolution). Figure \ref{ns_con_c}
shows the convergence of $\langle v_y \rangle$, the vertical
component of the velocity averaged over a spherical shell as a
function of the radius $r=\sqrt {x^2+y^2+z^2}$ of that shell, at
$t=8$ for each of these three resolutions.  We expect the vertical
velocity to be a good indicator of convergence since it measures the
buoyant motion of the bubble.  Note the overall radial profile of
the vertical velocity is very close at each resolution.  The
difference between the highest and our fiducial resolution
simulation is less than $10\%$ at every radius, and this difference
is everywhere smaller than that between the fiducial and the lowest
resolution simulations, which is the hallmark of convergence.

Figure \ref{ns_con_c} also shows the time evolution of the volume
averaged kinetic energy $E_k=\int_V \rho v^2/2 dV$ for each
different resolution.  Initially $E_k=0$, but as the bubble rises
the kinetic energy quickly evolves.  The largest difference in $E_k$
between the low and fiducial resolution runs is reached at $t=8$,
and is about $15\%$, while the difference between the standard and
high resolution cases is much smaller, about $5 \%$ at $t=8$. Once
again, the decrease in this difference with numerical resolution is
an indication of convergence.

\begin{figure}
  \includegraphics[width=150mm]{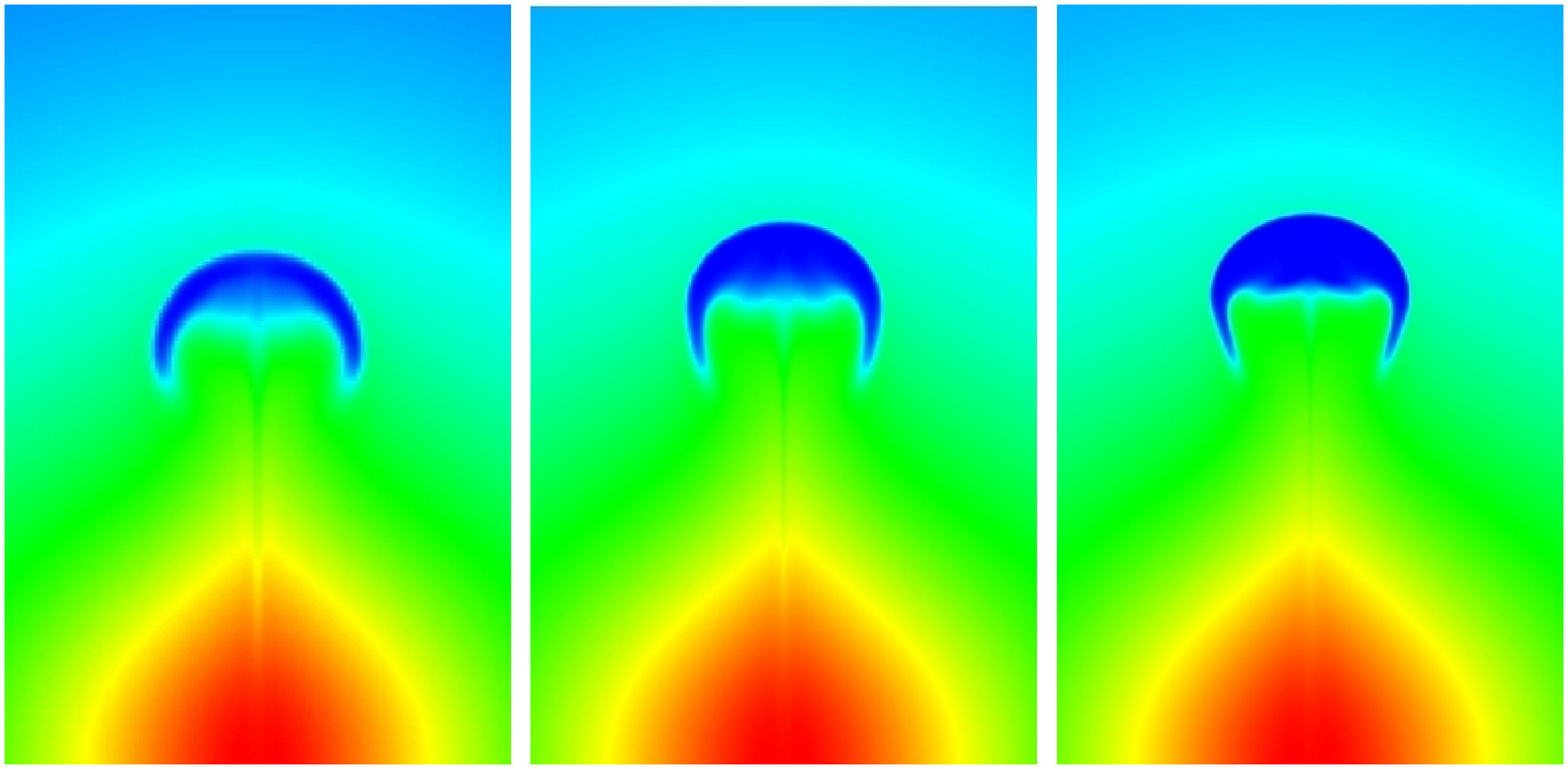}\\
  \caption{Slice of the density through the midplane ($z=0$) in run H2 at $t=8$
for three different numerical resolutions.
  From left to right: $128\times192\times128$, $256\times384\times256$, and
  $512\times768\times512$. There is very little difference in the structure
of the bubble between the latter two.  A linear color scale from
$\rho=0$ to $\rho =1$ is used.}
  \label{ns_con8}
\end{figure}

Finally,  Figure \ref{ns_con8} show the density of the bubble in a
slice through the midplane ($z=0$) in run H2 at $t=8$ using these three
numerical resolutions, in order to compare the morphology of the bubble
as an indication of convergence.  Note the bubble in the standard and
high resolution simulations is nearly identical, while the differences
between the standard and low resolution simulations are also small.
Based on these results, we have confidence that the standard resolution
used in this study is sufficient.

\subsection{Hydrodynamic Models}

\begin{figure}
  \includegraphics[width=160mm]{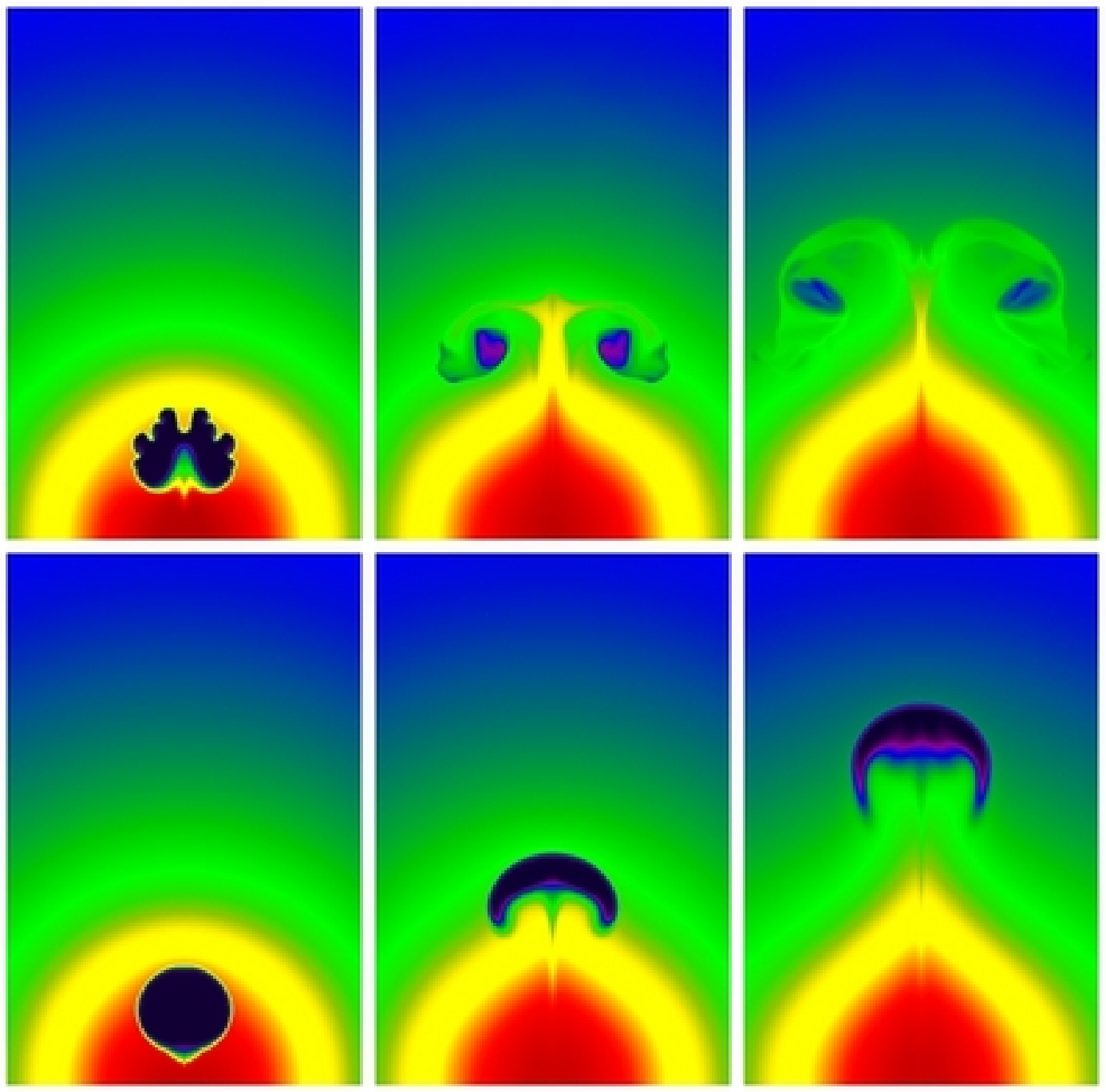}\\
  \caption{Slices of the density at the midplane ($z=0$) at $t=1$
  (left column), $t=4$ (middle column) and $t=8$ (right column) for
run H1 (top row) and H2 (bottom row). A linear color scale from
$\rho=0$ to $\rho =1$ is used.}
  \label{hydro_ns_density}
\end{figure}

To confirm that we reproduce previous inviscid and viscous hydrodynamic
simulations, and to provide a benchmark for comparison of the MHD models
presented below, we first discuss runs H1 (inviscid) and H2
(isotropic viscosity with $Re=50$).
Figure \ref{hydro_ns_density} shows slices of the density at the mid-plane
($z=0$) for these runs at three different times.  The most striking
feature in the plots is the emergence of RTI at the top of the bubble,
and KHI along the edges, in run H1 (top row of plots).  As a results of
these instabilities, the bubble interface is highly distorted, material
from the ICM can penetrate the bubble, and the bubble is subsequently
shredded and diffused into the ICM.  As shown by Reynolds et al. (2005),
an isotropic Navier-Stokes viscosity suppresses these instabilities, and
mixing.  The bottom row of plots shows run H2, which has a Reynolds number
of 50.  Clearly the evolution of the bubble is dramatically altered.
The viscosity quenches the instabilities, and the bubble remains intact
throughout the evolution.

Reynolds et al. (2005) also presented synthetic x-ray images of their
simulations, computed assuming an x-ray emissivity proportional to
$\rho^2 T^{0.5}$ which is then integrated along the line of sight.
These authors showed that the inviscid model (equivalent to H1) cannot
reproduce observed features such as cavities, since the bubble is shredded
by instabilities.  On the other hand, the morphology of the bubble in the
viscous case (equivalent to H2) shows good agreement with observations.
We have computed synthetic images of our simulations, and are able to
reproduce these results (see \S4.1).

\subsection{Uniform horizontal field}

\begin{figure}
  \includegraphics[width=160mm]{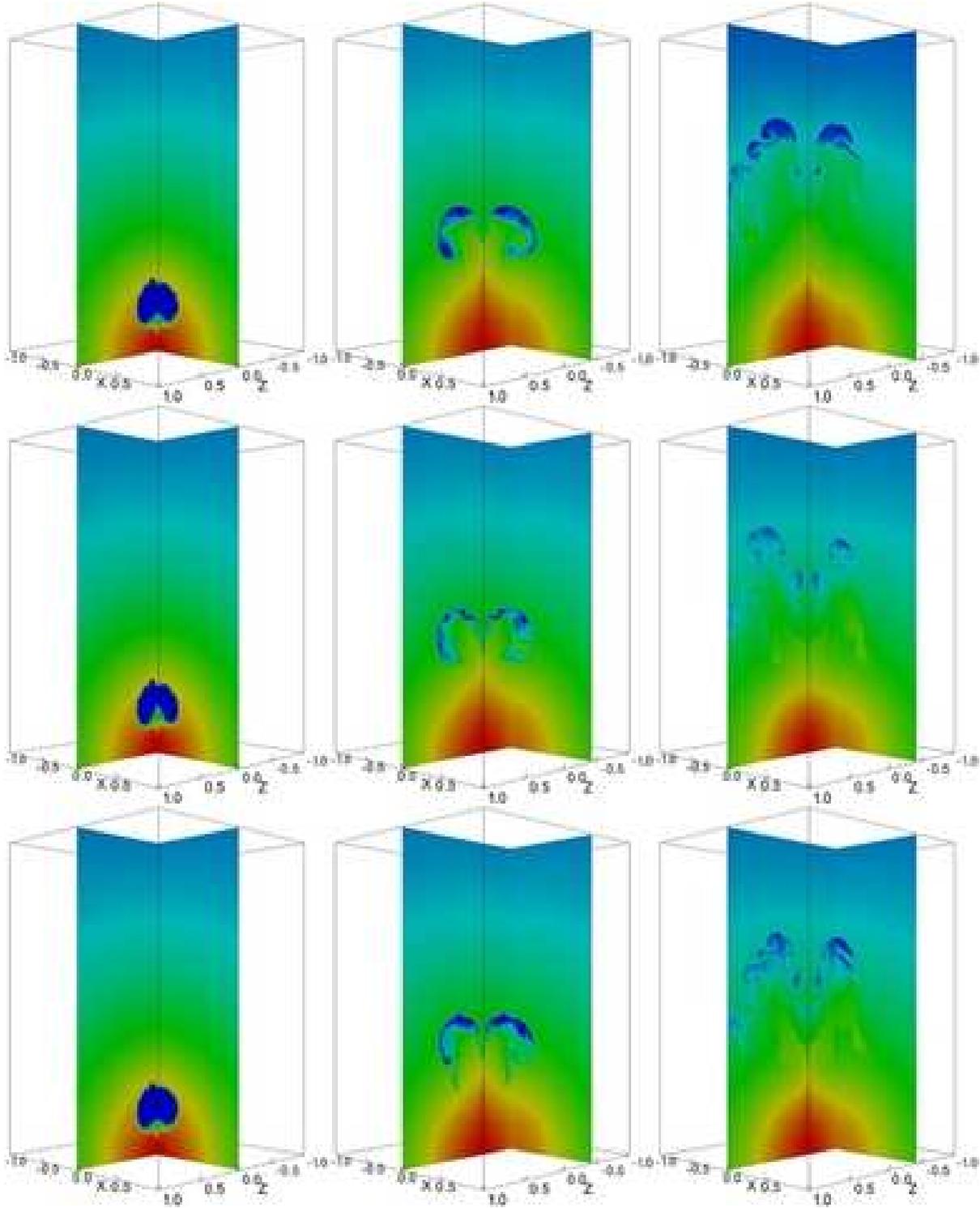}\\
\caption{Slices of the density in the $y-z$ plane (left half of each
plot) and $x-y$ plane (right half of each plot) at $t=1$ (left
column), $t=4$ (middle column) and $t=8$ (right column) for runs
beginning with an initially horizontal magnetic field.  The top row
shows the evolution of run X1, the middle row X2, and the bottom row
X3. A linear color scale from $\rho=0$ to $\rho=1$ is used. (Note
the $x-y$ plane plot at $t=1$ is at $z=0$, while at $t=4$ and $t=8$
it is at $z=-0.4$)} \label{Bx_density}
\end{figure}

We now turn to models with a uniform horizontal magnetic field throughout
the domain (including the bubble) initially.  Figure \ref{Bx_density}
shows slices of the density in runs X1 (weak horizontal field and
anisotropic viscosity), X2 (strong horizontal field and no viscosity),
and X3 (strong horizontal field and anisotropic viscosity), at three
different times, $t=1,4$ and 8.  Since the horizontal field breaks
axisymmetry, slices along both the planes $x=0$
and $z=0$ are shown using a perspective plot.  Initially, the field is
entirely in the $x-$direction, that is perpendicular to the slice at
$x=0$, and in the plane of the slice at $z=0$.

The top row of images shows the evolution of the density in model X1.
It is useful to compare this evolution directly to run H2 shown in the
bottom row in figure \ref{hydro_ns_density}.  Unlike the hydrodynamic
case with isotropic viscosity, the images show that in a weak magnetic
field with anisotropic viscosity, the bubble is shredded by RTI and KHI,
although in a complex manner.  In the $y-z$ plane, perpendicular to
the direction of the magnetic field, the bubble splits into two halves,
and there are clear KH rolls that develop in this plane.  In the $x-y$
plane, parallel to the field, the bubble surface shows less instability
and distortions.  Viscosity along the field lines in this plane suppresses
instability.

The second row of Figure \ref{Bx_density} shows the evolution of the
density in model X2.  Interesting, although there is no viscosity in
this calculation, the evolution is similar to model X1.  Once again,
the bubble is shredded by KHI and RTI, splitting apart with strong
mixing in the $y-z$ plane, and remaining more coherent in the $x-y$
plane. This evolution is a clear indication of the anisotropic
suppression of RTI in strong magnetic fields (Stone \& Gardiner
2007a; b).  Perpendicular to the field, interchange modes grow
rapidly, splitting the bubble and causing mixing, while parallel to
the field magnetic tension suppresses the RTI and the bubble remains
more intact.  In comparison to run X1, there is more diffusion and
mixing, and smaller scale structures develop in run X2.

Finally, the bottom row in Figure \ref{Bx_density} shows the
evolution of the density in model X3.  Once again, the overall
evolution is similar to runs X1 and X2: the bubble is split into two
halves, with strong mixing and small scale KHI in the $y-z$ plane,
and less instability and mixing in the  $x-y$ plane.  While in run
X1 (top row) the bubble remained intact until the final time, in run
X3 instabilities almost entirely destroy the bubble, with the hot
plasma in the bubble completely mixed with its surrounding colder
gas.  The similar evolution observed in all three of these runs is
an indication that uniform horizontal fields play a similar role to
anisotropic viscosity along weak fields of the same geometry in the
evolution of rising bubbles.

For the strong field runs X2 and X3, the final magnetic field geometry
largely maintains its initial structure at the final time, while in
the weak field run X1 the final field structure is highly tangled.
The ratio of the magnetic energies in the vertical and horizontal components
of the field $B_{y}^2/B_{x}^2$ is 0.76 for run X1,
0.32 for X2 and 0.23 for X3.  Thus, the field is still mostly horizontal
at the end of runs X2 and X2, while it is nearly isotropic in run X1.

\subsection{Uniform vertical field}

\begin{figure}
  \includegraphics[width=120mm]{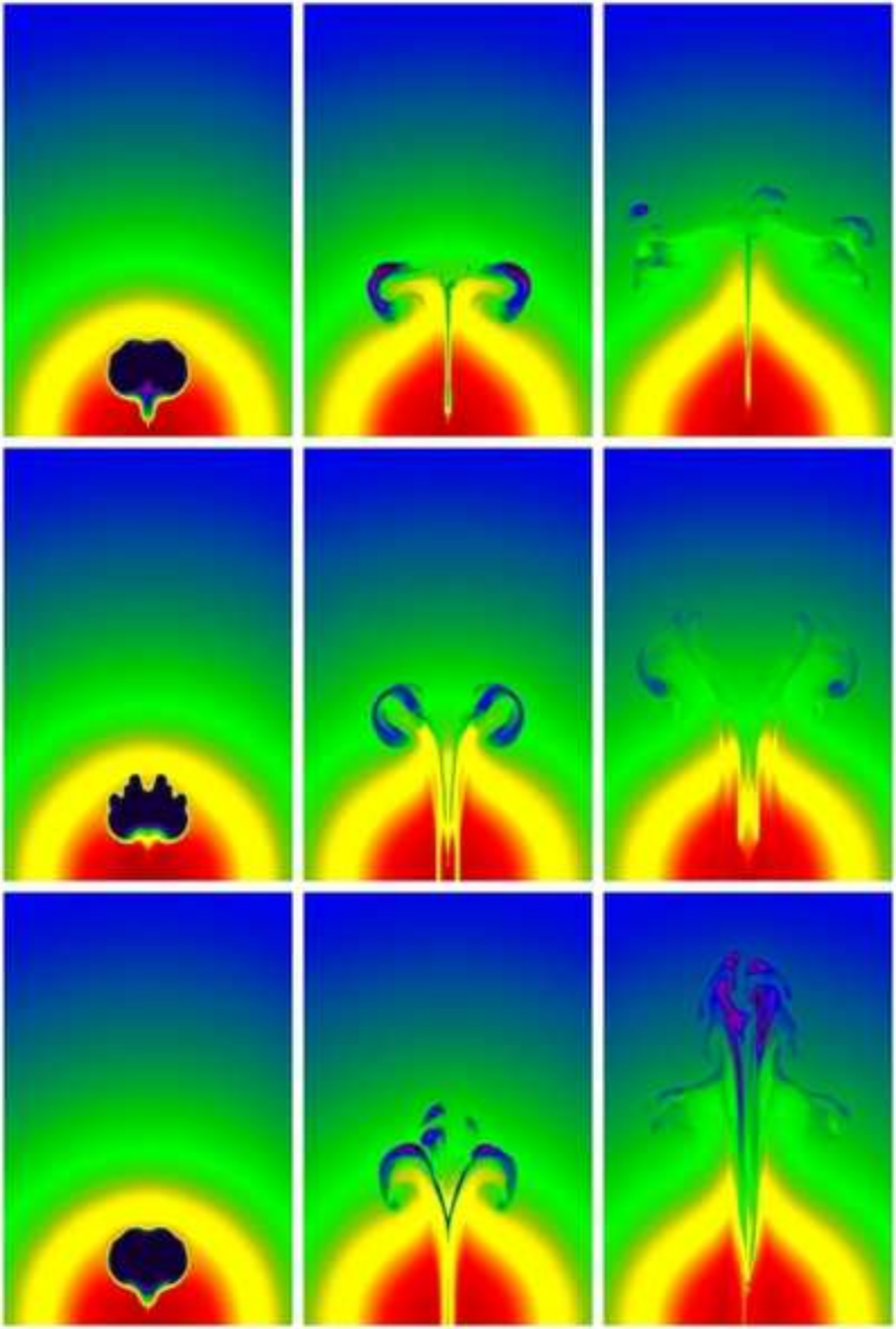}\\
  \caption{Slices of the density in the $x-y$ plane at $t=1$ (left
column), $t=4$ (middle column) and $t=8$ (right column) for runs
beginning with an initially vertical magnetic field.  The top row
shows the evolution of run Y1, the middle row Y2, and the bottom row
Y3. A linear color scale from $\rho=0$ to $\rho=1$ is used.}

\label{By_density}
\end{figure}

Figure \ref{By_density} shows slices of the density at three
different times ($t=1,4$, and 8) in the $x-y$ plane in runs Y1 (weak
vertical field and anisotropic viscosity), Y2 (strong vertical field
and no viscosity), and Y3 (strong vertical field and anisotropic
viscosity). Since the evolution is axisymmetric with vertical fields
apart from grid effects (with a Cartesian grid, the bubble actually
shows quadripole symmetry at late times), only one slice in any
vertical plane is sufficient to show the structure of the bubble.

The top row of images shows the evolution of run Y1.  In these runs,
the growth of secondary KHI produces vortex rolls that wind up the
vertical field.  In turn, the field tries to resist the winding
motion. In the weak field case shown in run Y1, the KHI largely
overwhelms the magnetic tension, and the bubble is shredded and
diffuses significantly into the ambient medium by the end of the
simulations, with little hot plasma still being inside the bubble.
Compared to the purely hydrodynamic model H2, we see that
anisotropic viscosity has altered the overall evolution; the bubble
splits into a ring instead of remaining coherent, and this ring
spreads laterally away from the axis of symmetry.  However, in
contrast to the hydrodynamic case, anisotropic viscosity in run Y1
does not suppress the shredding or mixing of the bubble.

The middle row in Figure \ref{By_density} shows the evolution in run
Y2.  As in run Y1, the bubble develops into a ring, with significant
lateral spreading.  The development of RTI is obvious at early
times, (the first image at $t=1$).  Large amplitude fingers are
clearly visible; the largest finger at the center of the bubble
ultimately results in splitting the bubble into a ring.  Note the
development of these fingers relies somewhat on the high degree of
symmetry in the initial conditions. If the bubble were not initially
axisymmetric, the resulting structure of the bubble would change.
Once again, the similarities between runs Y1 and Y2 indicate that a
strong field has similar effects on the dynamics of rising bubbles
as an anisotropic viscosity.

The bottom row in Figure \ref{By_density} shows the evolution in run Y3.
The evolution is quite different in comparison to the previous two.
Now, horizontal motions are strongly suppressed compared to vertical
motions, and a large amount of material originally in the bubble remains
concentrated within a small volume near the axis of symmetry at the end
of the simulation.  The initial spherical bubble evolves into a cone that
roughly occupies a vertical column with transverse width equal to the
original size of the bubble.  The bubble does not remain a single coherent
entity, however, but is split into many small columns by RTI fingers.

For these vertical field runs, the magnetic field geometry does not change
significantly with time, especially in the strong field with viscosity
case. The ratio of magnetic energies in the horizontal and vertical
components of the field $(B_{x}^2+B_{z}^2)/B_{y}^2$ is
0.10 for model Y1, 0.07 for Y2 and 0.03 for Y3.  This indicates for all
three runs that only a small amount of magnetic energy is transformed
from the vertical to the horizontal directions.  The field lines are
only weakly tangled, indicating the field roughly maintains its initial
ordered configuration.

To investigate the evolution of very strong vertical fields, we ran
several simulations with different values of $\beta \leqslant 1$.
Interestingly, we found the bubble does not rise at all in this case.
The upward movement of the bubble requires the ambient plasma above the
bubble to move laterally and around the sides.  With a very strong field,
the magnetic tension is so strong it prevents any lateral movement of the
fluid above the bubble, and the bubble is unable to rise.  We note this
result is quite different that reported by Robinson et al. (2004), who
found rising bubbles even with $\beta = 0.019$.  We are uncertain as to
the cause of this discrepancy, but speculate that it might
reflect differences in the numerical algorithms.

\subsection{Toroidal field confined to the bubble interior}

\begin{figure}
  \includegraphics[width=120mm]{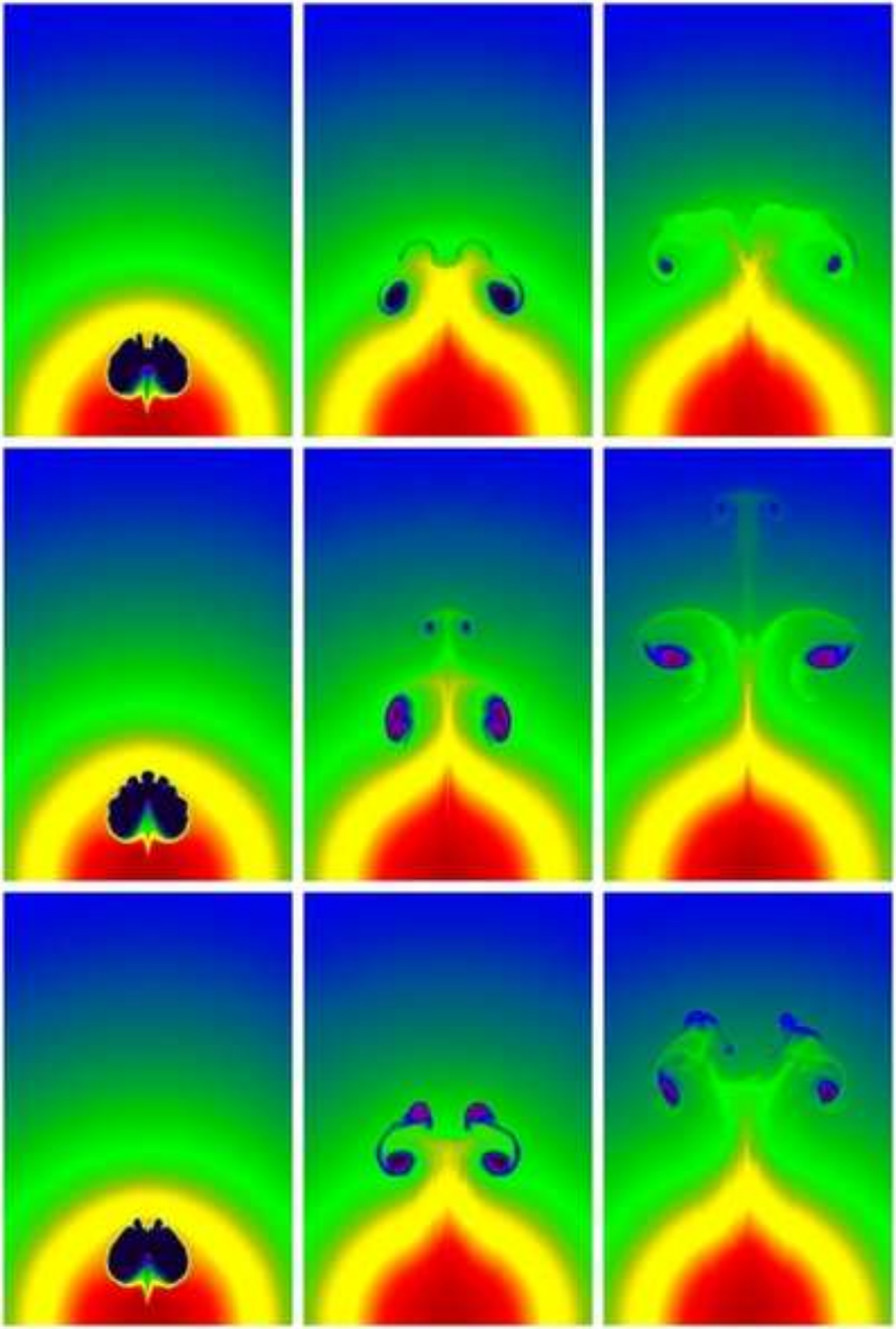}\\
  \caption{Slices of the density in the $x-y$ plane at $t=1$ (left
column), $t=4$ (middle column) and $t=8$ (right column) for runs
beginning with an initially toroidal magnetic field confined to the
interior of the bubble.  The top row shows the evolution of run T1,
the middle row T2, and the bottom row T3. A linear color scale from
$\rho=0$ to $\rho=1$ is used.}
  \label{Bt_density}
\end{figure}

Figure \ref{Bt_density} shows slices of the density  at three
different times ($t=1,4$, and 8) in the $x-y$ plane in runs T1 (weak
toroidal field just inside the bubble and anisotropic viscosity), T2
(strong toroidal field just inside the bubble and no viscosity), and
T3 (strong toroidal field just inside the bubble and anisotropic
viscosity).  As in the vertical field case, since the bubble
evolution is axisymmetric only one slice in a single vertical plane
is sufficient to show the structure of the bubble (as before, the bubble
actually has quadripole symmetry on our Cartesian grid).

The top row of images shows the evolution of run T1. As in the
vertical field case run Y1, the bubble evolves into a ring at late
times.  However, unlike run Y1, this ring remains relatively
coherent, and is not shredded by RTI and KHI.  Since the magnetic
field is everywhere parallel to the surface of the bubble, viscous
transport along the field lines is able to suppress significant
mixing with a toroidal field.

The middle row in Figure \ref{Bt_density} shows the evolution in run
T2. Once again due to the development of RTI fingers at early times,
the bubble forms a ring (actually two rings, with one small and
narrow ring formed above the main structure).  However, hoop
stresses associated with the strong field in this case keeps this
ring coherent, and even at late time the ring is quite prominent and
is not shredded by secondary KHI. Even at the end of the simulation
$t=8$, the ring contains a significant amount of hot plasma.

The bottom row in Figure \ref{Bt_density} shows the evolution in run
T3. Again, as in the previous two toroidal field simulation, the
bubble evolves into a ring.  At the earliest time ($t=1$), the
combination of viscosity and a strong field in run T3 suppresses the
RTI in comparison to run T2 -- there are fewer fingers in this case,
and they are smaller in comparison to run T2.  Hover, at late times
($t=8$), there is more small scale structure in the bubble in run T3
than T2, although the gross features of the ring are very similar in
both cases.  This demonstrates the complexity of the effects of
anisotropic viscosity.

\section{Discussion}

\subsection{Synthetic x-ray images}

One of the primary goals of this project is to study whether bubble
rising in a stratified atmosphere with realistic physics for the ICM
(magnetic fields and anisotropic viscosity) can produce structures
reminiscent of the cavities and filaments seen in the x-ray surface
brightness distribution of observed systems (e.g. Birzan et al.
2004).  To compute synthetic x-ray images of our simulations, we
assume an x-ray emissivity proportional to $\rho^2 T^{0.5}$, and
integrate along the line of sight.  As discussed in \S3.2, synthetic
x-ray images of our hydrodynamic simulations (both H1 and H2) look
identical to those presented in figure 4 of Reynolds et al. (2005),
and so we do not reproduce them here.  An important conclusion of
these authors is that inviscid evolution of the bubble shreds it
into small scale features that do not reproduce the observations. We
note that the X-ray surface brightness images generated from our
simulations are idealized, and do not contain noise or background
signal. Furthermore, the resolution of our synthetic images is the
spatial resolution used in the code, which typically is much higher
than the resolution in real observations. (For a computational
domain of size $40\times40\times60$ kpc, our image resolution is
0.16 kpc. For typical clusters like Cygnus A, about 237 Mpc from us
and Hydra A, about 240 Mpc form us, the image resolution on the
Chandra ACIS chips will be 0.5 kpc; while for closer clusters like
Perseus, about 72 Mpc from us, the resolution is comparable, about
0.17 kpc.)  For these reasons, detailed comparison between our
results and specific observations could be misleading, so that we
will focus only on the general characteristics of the images.

\begin{figure}
  \includegraphics[width=80mm]{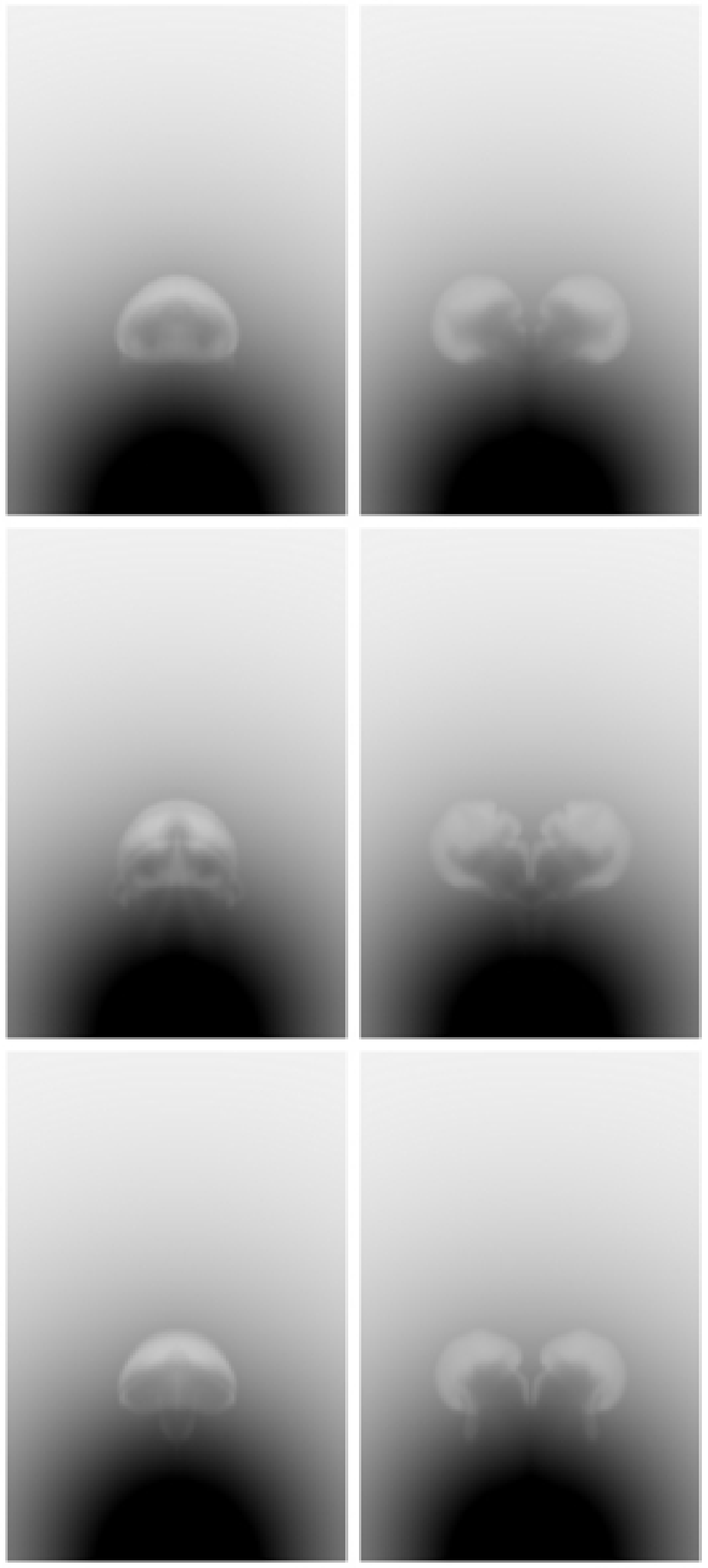}\\
\caption{Synthetic x-ray surface brightness images of runs X1 (top
row), X2 (middle row), and X3 (bottom row) at $t=4$.  Images in the
left column correspond to viewing the bubble along the $z-$axis, the
right column corresponds to the view along the $x-$axis.}
\label{Bx_x-ray}
\end{figure}

Figure \ref{Bx_x-ray} shows simulated x-ray surface brightness
images for runs X1, X2, and X3 at $t=4$, along two sightlines.  As
one might expect given the anisotropic structure of the bubbles
evident in Figure \ref{Bx_density}, the morphology of the bubble in
these images strongly depends in the sightline. When viewed
perpendicular to the $x-y$ plane, all three cases produced coherent
bubbles with a similar morphology: a bright cap with a darker cavity
behind.  This morphology is reminiscent of the structures of
hydrodynamic bubbles with viscosity (e.g. run H2; see figure 4 in
Reynolds et al. 2005), but shows more small scale structure in the
region below the cap. On the other hand, when viewed perpendicular
to the $y-z$ plane, the bubble is split into two crescent-shaped
bright structures, with a dark lobe below each.

It is clear from inspection of  Figure \ref{By_density}
that with vertical fields, the bubble is shredded by instabilities
even with anisotropic viscosity.  Thus, simulated x-ray images of
the bubbles with vertical fields do not show cavities or filaments
reminiscent of observations, and we do not show such images here.

\begin{figure}
  \includegraphics[width=120mm]{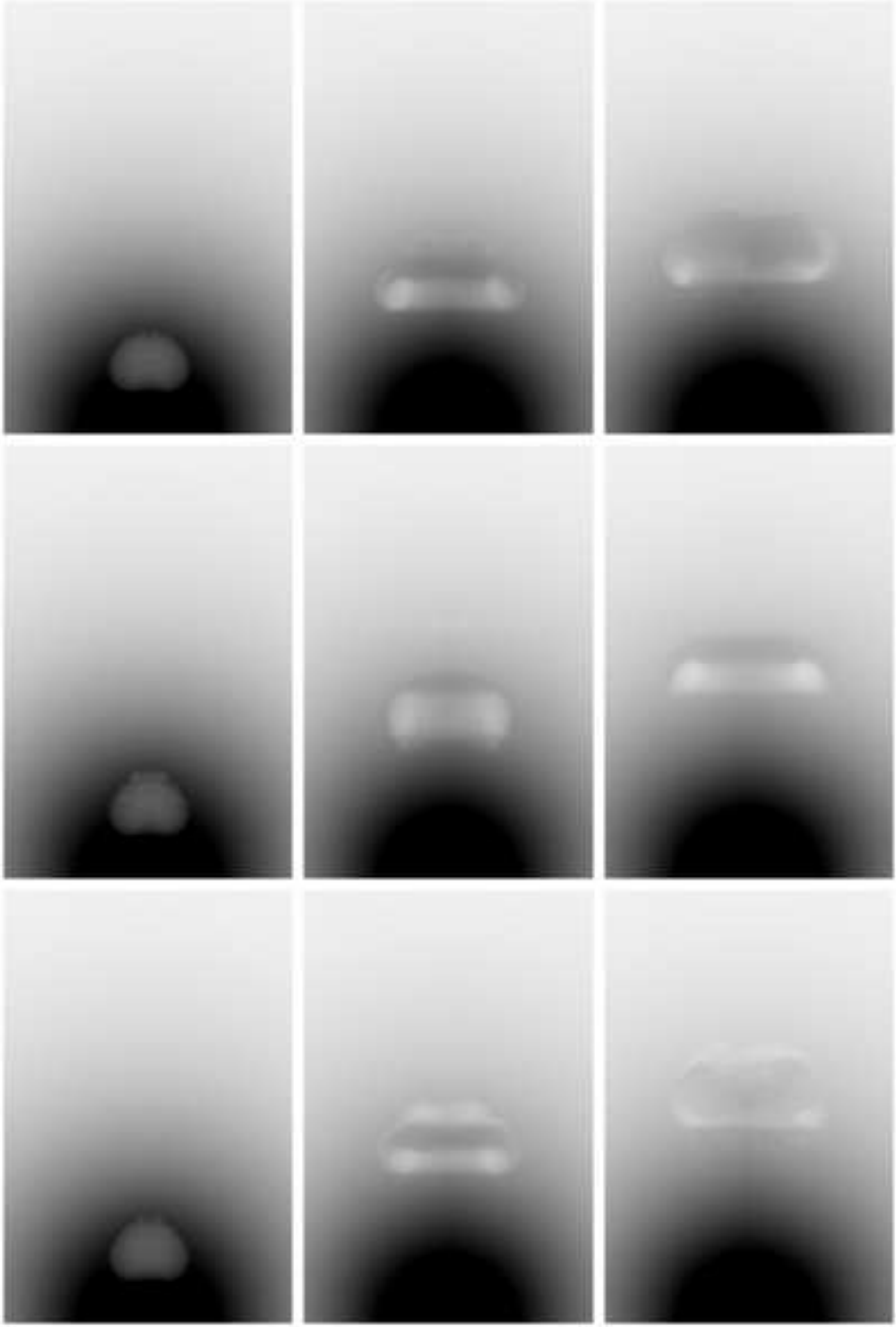}\\
  \caption{Synthetic x-ray surface brightness images for runs T1 (top row),
T2 (middle row), and T3 (bottom row) at $t=1$ (left column), $t=4$
(middle column), and $t=8$ (left column).}
  \label{Bt_x-ray}
\end{figure}

In Figure \ref{Bt_x-ray}, we show synthetic x-ray images of each of
the three toroidal field simulations, at three different times.  The
ring structure of the bubble produced in all these cases is obvious.
A toroidal field, either with or without viscosity, produces
cavities in the x-ray surface brightness similar to observations,
especially in runs T1 and T2 (top and middle rows of Figure
\ref{Bt_x-ray}). Although the structure formed here is ring-like
instead of a uniform bubble, it is impossible to rule out this
morphology given current observations. The image quality and
resolution of current X-ray telescopes do not allow distinction
between the two structures based on their two-dimensional surface
brightness.

We conclude that initially horizontal fields (when viewed
perpendicular to the direction of the field), or toroidal magnetic
fields confined to the interior of the bubble, produce ``ghost
cavity" structures reminiscent of x-ray images by Chandra and
XMM-Newton. However, for vertical fields, such cavities cannot be
reproduced.  Both strong magnetic fields, or weak fields with
anisotropic viscosity, are able to reproduce these features with the
appropriate initial field geometry.  The most realistic field
geometry for bubbles inflated by AGN is likely a toroidal field
confined to the interior of the bubble.  Thus, the fact this
geometry reproduces observations is reassuring.

\subsection{Heating and AGN feedback}

\begin{figure}
  \includegraphics[width=160mm]{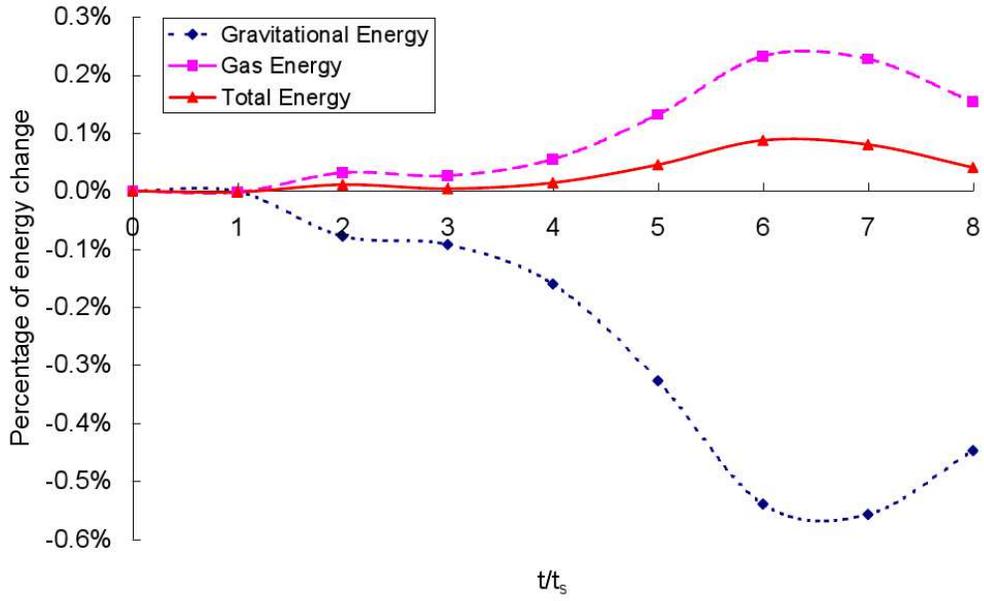}\\
  \caption{Time evolution of the volume averaged gravitational
energy $E_G$, total energy $E_T$, and the gas energy, which is the
difference between the two $\delta E = E_t - E_G$, for run H1.}
  \label{hydro_severalE}
\end{figure}

Another important goal of this study is to attempt to measure the
heating rate of the ICM due to MHD processes associated with rising
bubbles generated by AGN, using more realistic physics (magnetic
fields and anisotropic viscosity). However, because of our use of
open (outflow) boundary conditions in the simulations, total energy
is not conserved, which can complicate such measurements. To
investigate how much the total energy changes, compared to the
changes in the internal energy, Figure \ref{hydro_severalE} shows
the fractional change in the volume averaged gravitational energy
($E_G= \int \rho \phi dV/V$), the volume averaged total energy $E_t$
(sum of the kinetic, internal, and gravitational energies) and the
difference between the two $\delta E = E_t - E_G$, scaled by their
initial values, as a function of time $t$ in run H1. The
gravitational energy first decreases as the bubble rises due to
buoyancy. This process releases potential energy which can be
converted into other forms.  The total energy $E_t$ increases a
little with time, reaching a maximum at about $t=6$, at which point
it is only $0.1 \%$ larger than the initial value. This increase is
due to work done on the system at the boundary. However, note that
the change in total energy is less than the gravitational potential
energy released by the rising bubble.  Thus, any increase in the
other forms of energy, including the internal energy of the ICM,
will be dominated by the release of gravitational energy, and not
the work done on the boundaries. Thus, tracking the changes in each
form of energy in our simulations should give us a good estimate of
the energy input due to the dynamics of the rising bubble.

\begin{figure}
  \includegraphics[width=160mm]{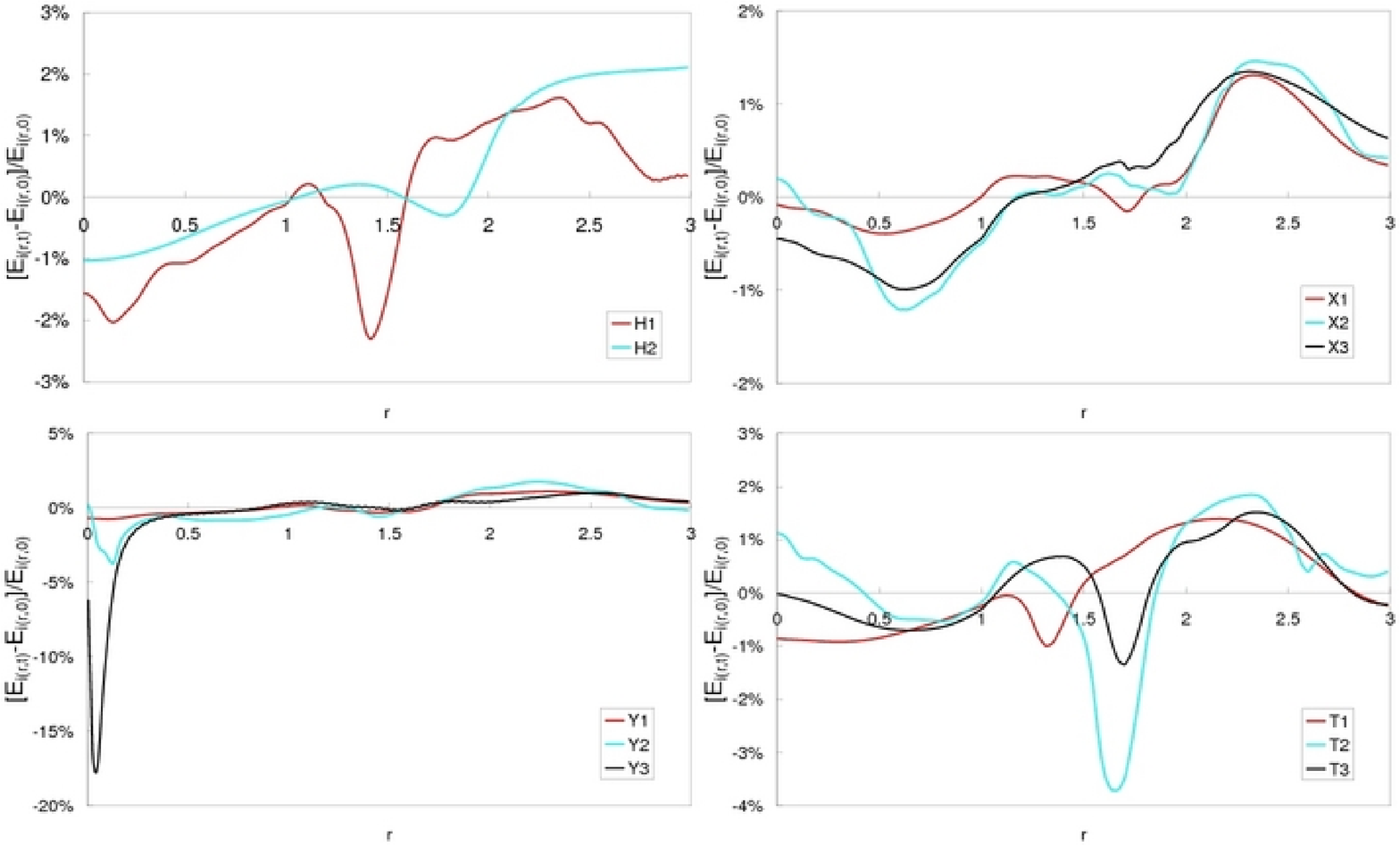}\\
  \caption{Fractional change in internal energy $(E_{i(r,t)}-E_{i(r,0)})/E_{i(r,0)}$
as a function of spherical radius from the center of the cluster
between final and initial states, scaled by the corresponding
initial internal energy for all 9 MHD and two hydrodynamical runs.
Panels are (from right to left, and from top to bottom):
hydrodynamic runs H1 and H2, horizontal field runs X1, X2, and X3,
vertical field runs Y1, Y2, and Y3, and toroidal field runs T1, T2,
and T3.}
  \label{ET_r}
\end{figure}

Figure \ref{ET_r} shows the fractional change of volume averaged
internal energy $E_i = E_t - E_k - E_B - E_G$ (where $E_t$, $E_k$,
$E_B$ and $E_G$ are the volume averaged total, kinetic, magnetic and
gravitational energies respectively) as a function of spherical
radius $r$ between the final (at $t=8$) and initial states for all 9
MHD and two hydrodynamic simulations presented in this paper.
Regions where this change is positive indicate net heating between
the initial and final state, while regions where the curve is
negative indicate net cooling between the initial and final state.
Note here that changes in the internal energy are not necessarily
equivalent to heating or cooling, but they are an indication that
such processes might be occurring. The curves in the panels
corresponding to the hydrodynamic, horizontal field, and toroidal
field simulations are all qualitatively similar. Net heating of
between 1\% and 2\% occurs at large radii, beyond $r=2$ (or 40 kpc
for our fiducial parameters). The heating region corresponds to
location of the bubble at the end of the simulation, and in fact the
increase in internal energy is largely a reflection of the
displacement of the background ICM by the hot plasma in the bubble.
Interior to region of heating there can be strong cooling,
especially in the toroidal field case. This is likely caused by
adiabatic expansion of the plasma immediately below the bubble as it
rises. Generally, internal energy is lost throughout the region
below the bubble.

The panel corresponding to the vertical field cases also show an
increase in the internal energy of a few percent at the location of
the bubble, but they also show strong cooling at the center, at
least for the strong field cases.  This is likely caused by
adiabatic expansion of the plasma below the bubble as it rises.
With a vertical field, transverse motions that can replace the
material below the bubble are suppressed, which prevents transverse
compressional heating that can replace the vertical expansion and
associated cooling.

It is difficult to assess the implications of these results on the
long term thermal stability of the ICM.  Our simulations follow the
dynamical evolution of one bubble over roughly 100~Myr, rather than an
ensemble of bubbles produced by jet activity over cosmological
timescales.  Nevertheless, our results show that the most
realistic field geometries (e.g. toroidal field within the bubble)
produces a net heating of a few percent in the outer regions.  None
of our simulations produces strong heating in the inner regions,
where the cooling times are shortest.

\section{Conclusions}

We have studied the buoyant rise of bubble including the effect of
magnetic fields and anisotropic (Braginskii) viscosity. Our primary
conclusions are the following:

1. Both strong magnetic fields and anisotropic viscosity have similar
effects on the evolution of rising bubbles. They both can suppress
instabilities in the direction parallel to the field,
while having little effect on interchange instabilities that develop
perpendicular to the field.

2. The detailed evolution of the bubble depends on the initial field direction.
A toroidal field confined to the bubble interior produces structures
that are most consistent with observed cavities and filaments.

3. Rising bubbles do not dramatically alter the initial structure of strong
magnetic fields.  In particular, strong horizontal fields remain mostly
horizontal even after the passage of the bubble, as measured by the ratio
of the magnetic energy in the vertical as compared to the horizontal
component of the field.

4. All the models we have run show an increase in the internal energy
(heating) of a few percent in the outer regions at late time, and a
decrease in the internal energy (cooling) in the region below the bubble.
Our simulations do not show that buoyant bubbles are an effective
mechanism for heating the ICM in the central regions of the cluster.

The primary limitation of our study is that we neglect several physical
processes that are likely to be important in real clusters.  We do not
follow the inflation of the bubble by an AGN self-consistently, rather
we study the buoyant evolution of a bubble initially at rest in the ICM.
It is possible that the transient evolution associated with the initial
upward acceleration of the bubble will affect the resulting stability
and mixing, thus it is more realistic to follow the inflation of the
bubble directly (Pizzolato \& Soker 2006).  Furthermore, we study mostly
uniform magnetic fields, whereas the field in real clusters is likely
to be highly tangled.  Finally, we have neglected the effect of thermal
conduction, cosmic rays, and turbulent motions in the ICM driven
by merger of substructures.  Nevertheless, our idealized models allow
us to isolate the effect of specific physics, namely magnetic fields and
anisotropic viscosity, on the dynamics of the ICM.  We find that both have
a significant effect, and should be included in more realistic studies.

In the future, it would be productive to study the interaction of
jets with the ICM directly.  It is likely that adaptive mesh refinement
(AMR) simulations that can simultaneously resolve the jet and global
cluster structure will be required.  Our results indicate it will
be necessary to include the effects of MHD and anisotropic transport
coefficients in such models to capture the physics of the ICM correctly.

\section*{Acknowledgments}

We thank W. Mathews and F. Brighenti for useful conversations. We
also acknowledge useful suggestions from Chris Reynolds, and an
anonymous referee. Simulations were performed on system supported by
the Princeton Institute for Computational Science and Engineering,
and NSF grand AST-0722479.

\end{document}